\documentclass{article}
\usepackage{graphicx}
\begin{document}
\title{Sum rules for total hadronic widths of light mesons and  
rectilineal stitch of the masses on the complex plane}
\author{Micha{\l} Majewski} \maketitle

\begin{abstract}
  Mass formulae for light meson multiplets derived by means of exotic
  commutator technique are written for complex masses and considered
  as complex mass sum rules (CMSR). The real parts of the (CMSR) give
  the well known mass formulae for real masses (Gell-Mann--Okubo,
  Schwinger and Ideal Mixing ones) and the imaginary parts of CMSR
  give appropriate sum rules for the total hadronic widths - width sum
  rules (WSR). Most of the observed meson nonets satisfy the Schwinger
  mass formula (S nonets). The CMSR predict for S nonet that the
  points $(m,\Gamma{})$ form the rectilinear stitch (RS) on the
  complex mass plane. For low-mass nonets WSR are strongly violated
  due to ``kinematical'' suppression of the particle decays, but the
  violation decreases as the mass icreases and disappears above $\sim
  1.5 GeV$. The slope $k_s$ of the RS is not predicted, but the data show
  that it is negative for all S nonets and its numerical values are
  concentrated in the vicinity of the value $-0.5$. If $k_s$ is known for
  a nonet, we can evaluate ``kinematical'' suppressions of its
  individual particles. The masses and the widths of the S nonet
  mesons submit to some rules of ordering which matter in
  understanding the properties of the nonet. We give the table of the S
  nonets indicating masses, widths, mass and width orderings. We show
  also mass-width diagrams for them. We suggest to recognize a few
  multiplets as degenerate octets. In Appendix we analyze the nonets
  of $1^+$ mesons.
\end{abstract}

\section{Introduction}
Total particle width is one of its main attributes just as 
important as mass and discrete quantum numbers. It tells us something
different than the mass and sometimes it may tell more. The widths of
the particles with similar masses may differ by many orders. Then the
widths first inform us which interaction -- strong, electromagnetic or
weak is responsible for their decay.

Obviously, the total hadronic widths are not so much differentiated,
but still are remarkably various.  Within $SU(3)_f$ meson multiplet
the differences are often of the same magnitude as between the masses.
Thus they merit attention. However, as the mass formulae have been
derived long ago, no relation is known between the total widths.
Perhaps such situation is due to conjecture that it is not worthy to
pay attention to the widths pattern of the multiplet, as the widths
are in a way accidental. Indeed, selection rules and phase space may
suppress more or less the decay of particular particle thus destroying
any given regularity. Such an effect should be especially transparent
in low-mass multiplets where e.g., for some particle two-body decays
are forbidden and many-body decays are suppressed (the meson $\omega$
is a striking example). For more massive multiplets, where many decay
channels are opened, we may expect better agreement. However, the
prediction may be interesting in any case.

\section{Exotic commutator mass sum rules (ECMSR)}
\label{sec:1}
The approach is based on the exotic commutator technique \cite{O}.  We
assume that the following set of exotic commutators vanishes (\cite{1})
\begin{equation}
\left[T_\alpha,\frac{d^jT_\beta}{dt^j}\right]=0,\quad   (j=1,2,3,...) \label{0.1}
\end{equation}
where the $T$ are $SU(3)_f$ generators, $t$ is time and $(\alpha,\beta)$
is an exotic combination of indices; that means that $T_\alpha$,
$T_\beta$ are chosen such that operator $[T_\alpha,T_\beta]$ does not
belong to the octet representation (we use the combination changing
strangeness by two). Putting
\begin{displaymath}
\frac{dT}{dt}=i[H,T]
\end{displaymath}
and using the infinite momentum approximation for the one-particle hamiltonian
\cite{1a}
\begin{displaymath}
H = \sqrt{m^2+p^2}\simeq p^2 +\frac{m^2}{2p} + O(\frac{1}{p^2}),
\end{displaymath}
we transform eqs.~(\ref{0.1}) into the following system (cf \cite{1})
\begin{displaymath}
[T_\alpha,[\hat{m^2},T_\beta]]=0,
\end{displaymath}
\begin{displaymath}
[T_\alpha,[\hat{m^2},[\hat{m^2},T_\beta]]]=0,
\end{displaymath}
\begin{equation}
[T_\alpha,[\hat{m^2},[\hat{m^2},[\hat{m^2},T_\beta]]]]=0, \label{0.2}
\end{equation}
\begin{displaymath}
........................................
\end{displaymath}
where $\hat{m^2}$ is the squared-mass operator. Taking matrix elements of these
operator equations between one-particle octet states and saturating the products 
of flavoured operators with one-particle intermediate octet states, we get
the system of linear equations for the matrix elements
$\langle{x_8}\mid{\hat{(m^2)}^j}\mid{x_8}\rangle$ $(j=1,2,3,...)$, where
$\mid{x_8}\rangle$ is the isoscalar octet state. Solving this system,
we find
\begin{equation}
\langle{x_8}\mid{\hat{(m^2)}^j}\mid{x_8}\rangle = \frac{1}{3}a^j + \frac{2}{3}b^j 
\quad (j=1,2,3,...). \label{0.3}
\end{equation}
Here $a$ is an isovector particle mass squared and
\begin{equation}
b=2K-a, \label{0.4}
\end{equation}
with $K$ being the mass squared of the isodublet particle. Notice that
\begin{equation}
\langle{x_8}\mid{\hat{m^2}}\mid{x_8}\rangle = \frac{1}{3}a + \frac{2}{3}b = x_8 \label{0.5}
\end{equation}
is the Gell-Mann--Okubo mass.

Let us discuss a nonet. Introduce isoscalar physical states
$|x_1\rangle$, $|x_2\rangle$ which are linear combinations of the
exact symmetry octet $|x_8\rangle$ and singlet $|x_0\rangle$ states.
Consequently, we can write
\begin{equation}
|x_8\rangle=l_1 |x_1\rangle +l_2 |x_2\rangle, \label{0.6}
\end{equation}
where $l_1$,$l_2$ are real and
\begin{equation}
l_1^2 + l_2^2 = 1. \label{0.7}
\end{equation}
Using (\ref{0.6}), we transform eqs.~(\ref{0.3}) into the system of
equations
\begin{equation}
l_1^2 z_1^j +l_2^2 z_2^j =\frac{1}{3} a^j +\frac{2}{3} b^j,\quad (j=1,2,3,...) \label{0.8}
\end{equation}
where $x_1,x_2$ are masses squared of the physical isoscalar mesons
(we choose $x_1< x_2$).

Eqs.~(\ref{0.7}) and (\ref{0.8}) are basic for the further investigation.
They are considered as linear equations for the unknown coefficients
$l_1^2$, $l_2^2$ and they are examined by the gradually enlarging number of
eqs.~(\ref{0.8}). This enables us to obtain by a uniform procedure the
known results of the broken $SU(3)_f$ symmetry. In the succeeding steps,
beginning from the first of eq.~(\ref{0.8}), we find the known mass
formulae: Gell-Mann -- Okubo (GMO), Schwinger (S) and Ideal Mixing
(IM). We also find the mixing angle in each of these cases. Let us
discuss the procedure in more detail.

If we consider only eq.~(\ref{0.7}) and the first of eq.~(\ref{0.8}) (one 
exotic commutator), we find
\begin{equation}
l_1^2=\frac{x_2-x_8}{x_2-x_1},\quad l_2^2=\frac{x_8-x_1}{x_2-x_1}.\label{0.9}
\end{equation}
Hence, for an octet ($l_1^2=1$, $l_2^2=0)$ we get GMO mass formula
\begin{equation}
x_1  = x_8. \label{0.10}
\end{equation}
For a nonet we have $l_1^2>0$, $l_2^2>0$. Introducing the mixing angle 
$\theta^{GMO}$ 
\begin{equation}
l_1^2 = \sin^2{\theta^{GMO}}, \label{0.11}
\end{equation}
we find
\begin{equation}
z_1\sin^2 \theta^{GMO} + z_2 \cos^2 \theta^{GMO} = z_8.\label{0.12}
\end{equation}
This equation determines $\theta^{GMO}$ which is usually quoted as the 
nonet mixing angle. There is no mass formula in this case and the 
$\theta^{GMO}$ is determined only by the experimental masses. Such a nonet
we call the GMO one.

If we consider eq.~(\ref{0.7}) and the first two of eqs.~(\ref{0.8}) 
(two exotic commutators),
then we have three equations for two unknown quantities. Solution 
(\ref{0.9}) remains true provided the masses satisfy a consistency
condition. This condition is the well known S mass formula. In terms of
the parameters $a$ and $b$ the mass formula and the coefficients $l_1^2$,
$l_2^2$ are:
\begin{equation}
(a-x_1)(a-x_2)+2(b-x_1)(b-x_2)=0, \label{0.13}
\end{equation}
\begin{equation}
l_1^2=\frac{1}{3}\frac{(x_2-a)+2(x_2-b)}{x_2-x_1}, \label{0.14}
\end{equation}
\begin{equation}
l_2^2=\frac{1}{3}\frac{(a-x_1)+2(b-x_1)}{x_2-x_1}. \label{0.15}
\end{equation}
These equations define another mixing angle, the $\theta^{Sch}$ one.
Such a nonet we call an S one.

If we consider three of eqs.~(\ref{0.8}) (three exotic comutators), 
then we get one more consistency condition
\begin{equation}
a(a-x_1)(a-x_2)+2b(b-x_1)(b-x_2)=0, \label{0.16}
\end{equation}
and consequently we obtain the IM nonet:
\begin{equation}
z_1 = a,\;\quad  z_2 = b \quad{\rm and}\quad     
l^{2}_{1} = \frac{1}{3},\;\quad   l^{2}_{2} = \frac{2}{3}.\label{0.17}
\end{equation}
The signs of $l_1$, $l_2$ are not determined.  Ideally mixed isoscalar
states arise, if $l_1l_2<0$.

It is now obvious that any additional eq.(\ref{0.8}) (for j=4,5,...)
is an identity.

{\em The exotic commutator approach is the only known way to derive the
ideally mixed states from something else. Alternatively they are
postulated.}

The current shape of the nonet mass formula
\begin{equation}
(x_2 - x_8)(x_8 - x_1) = \gamma^2,  \label{0.18}
\end{equation}
is obtained by diagonalizing the mass matrix
\begin{displaymath}
\left[
\begin{array}{lr}
x_0&\gamma\\
\gamma&x_8 
\end{array}
\right],
\end{displaymath}
and eliminating $x_0$, which is the mass squared of the $SU(3)_f$
singlet.  The GMO mass $x_8$ is known for all -- GMO, S and IM nonets.
The mixing parameter $\gamma$ is undetermined for the GMO nonet, but it is
known for the S nonet. It can be calculated from the mass formula
(\ref{0.13}) by the substitutions
\begin{equation}
a=x_8-\frac{2}{3}(b-a) \label{0.19},
\end{equation}
\begin{equation}
b=x_8+\frac{1}{3}(b-a) \label{0.20}
\end{equation}
and observation that
\begin{equation}
b-a=2(K-a) \label{0.21}.
\end{equation}
The calculation confirms ``Schwinger Ansatz'':
\begin{equation}
\gamma^2=\frac{8}{9}(K-a)^2. \label{0.22}
\end{equation}

{\em Data show (or at least suggest) that, with one exception of
  pseudoscalar mesons, all nonets comply with this expression.  So
  they are the S nonets.}

The pseudoscalar mesons $\pi{}$, $K$, $\eta{}$, $\eta'$ form the only
recognized GMO nonet. Its experimental masses are described by
eq.~(\ref{0.18}) with mixing parameter
\begin{equation}
\gamma^2=\frac{2}{9}(K-a)^2. \label{0.23}
\end{equation}

The S nonets are not much different from IM ones, but none of them is
strictly ideal (see Tab.~1 for an explicite comparision). 

For the IM nonet not only $\gamma^2$ is determined (eq.~(\ref{0.22})),
but also $x_0$ can be calculated:
\begin{equation}
x_0=\frac{1}{3}(2a+b). \label{0.24}
\end{equation}

\section{Sum rules for complex masses (CMSR)}
\label{sec:2}
\subsection{Complex mass operator}
\label{sec2.1}

We assume that eqs.~(\ref{0.8}) may be considered for complex
mass squared.  We choose the complex mass operator in the form
\begin{equation}
\hat{m_c^2}=\hat{m^2}-i\hat{m}\hat{\Gamma}, \label{3.1}
\end{equation}
where $\hat{m}$ and $\hat{\Gamma}$ are hermitean and commute. This
operator can be diagonalized and has orthogonal eigenfunctions. That
follows from the observation that the operators
\begin{equation}
\frac{1}{2}(\hat{m_c^2}+\hat{m_c^2}^+)=\hat{m^2}\quad{\rm and}
\quad\frac{i}{2}(\hat{m_c^2}-\hat{m_c^2}^+)=\hat{m}\hat{\Gamma}\label{3.2}
\end{equation}
are hermitean and commute.

We use the notation   
\begin{equation}
a_c=a-i\alpha,\;\quad  K_c=K-i\kappa,\;
\quad  z_1=x_1 -iy_1,\;\quad z_2=x_2-iy_2\;\label{3.3}
\end{equation}
for the complex masses of the physical particles, where
\begin{equation}
a=m_a^2,\quad K=m_K^2,\quad  x_1=m_1^2,\quad x_2=m_2^2\label{3.4}
\end{equation}
and 
\begin{equation}
\alpha=m_a\Gamma_a,\quad \kappa=m_K\Gamma_K,
\quad y_1=m_1\Gamma_1,\quad y_2=m_2\Gamma_2. \label{3.5}
\end{equation}
For subsidiary states with complex masses 
\begin{equation}
z_8=\frac{1}{3}a_c+\frac{2}{3}b_c,\;\quad b_c=2K_c-a_c    \label{3.6}
\end{equation}
we use a similar notation:
\begin{equation}
 z_8=x_8-iy_8,\;\quad  b_c=b-i\beta. \; \label{3.7}
\end{equation}

The parameters $\alpha$, $\kappa{}$, $y_1$, $y_2$ are positive, as
they refere to real particles. Below it will be seen that $y_8$ is
also positive. This concerns the observed values as well as the
predicted ones.  The question about positivity of $\beta{}$ is not so
simple. It will be seen further that the predicted value of $\beta{}$
is positive for all known S nonets, but as
\begin{equation}
\beta=2\kappa-\alpha, \label{3.7a}
\end{equation}
it may happen that $\beta<0$ for observed values of $\alpha$ and
$\kappa$. That will indicate a relative suppression of the $K$-meson 
decay. 
 
It is convenient to introduce the subsidiary widths $\Gamma{_8}$ and
$\Gamma{_b}$ :
\begin{equation}
y_8=m_8\Gamma_8,\; \quad \beta=m_b\Gamma _b. \label{3.8}
\end{equation}

For complex masses also the coefficients $l_1$, $l_2$ (\ref{0.6}) are
complex and in eqs.~(\ref{0.7}),(\ref{0.8}) the  $l^2$s are replaced
by the $\mid{l}\mid^2$s.

Below we show that the real parts of the masses (\ref{3.1}) obey the
usual mass formulae. For the imaginary parts there arise some new
relations.  We call them `` width sum rules'' (WSR) invoking the name
``mass sum rules'' (MSR) used sometimes for the mass formulae.

\subsection{Gell-Mann -- Okubo nonet}
\label{sec:2.2}

From eq.~(\ref{0.7}) and the first of eqs.~(\ref{0.8}) (for j=1) we find
\begin{equation}
\mid{l_{1}}\mid^{2}=\frac{x_{2}-x_{8}}{x_{2}-x_{1}}, \quad
\mid{l_{2}}\mid^{2}=\frac{x_{8}-x_{1}}{x_{2}-x_{1}}\label{3.9}
\end{equation}
and
\begin{equation}
\mid{l_{1}}\mid^{2}y_{1}+\mid{l_{2}}\mid^{2}y_{2}=y_{8}.\label{3.10}
\end{equation}
$\mid{l_{i}}\mid^{2}$ determines the mixing angle which depends only
on the masses and is not affected by the widths. Eq.~(\ref{3.10})
shows that $y_8$ is positive.

It is seen from (\ref{3.9}) that
\begin{equation}
x_{1}<x_{8}<x_{2}. \label{3.11}
\end{equation}
Therefore,
\begin{equation}
\tilde{x}_1=x_{8}-x_{1},\quad \tilde{x}_2=x_{2}-x_{8}\label{3.12}
\end{equation}
are positive. Introducing also 
\begin{equation}
\tilde{y}_1=y_{8}-y_{1},\quad \tilde{y}_2=y_2-y_8, \label{3.13}
\end{equation}
we can write (\ref{3.10}) in the form
\begin{equation}
\frac{\tilde{y}_2}{\tilde{x}_2}=\frac{\tilde{y}_1}{\tilde{x}_1}.\label{3.14}
\end{equation}

\subsection{Schwinger nonet}
\label{sec:2.3}
Consider eq.~(\ref{0.7}) and the first two of eqs.~(\ref{0.8}) (for j=1,2). 
Using for complex masses eq.~(\ref{0.18}) with $\gamma^2$ given by 
(\ref{0.22}), we find for its real part 
\begin{equation}
\tilde{x}_1\tilde{x}_2-\tilde{y}_1\tilde{y}_2
=\frac{2}{9}[(b-a)^2-(\beta -\alpha )^2]\label{3.15}
\end{equation}
and for the imaginary one 
\begin{equation}
\tilde{x}_1\tilde{y}_2+\tilde{x}_2\tilde{y}_1=
\frac{4}{9}(b-a)(\beta -\alpha ).\label{3.16}
\end{equation}
From eqs.~(\ref{3.14}) and (\ref{3.16}) we find
\begin{eqnarray}
\tilde{y}_1&=&\frac{2}{9}\frac{(b-a)(\beta -\alpha )}{\tilde{x}_2},\label{3.17}\\
\tilde{y}_2&=&\frac{2}{9}\frac{(b-a)(\beta -\alpha )}{\tilde{x}_1}.\label{3.18}
\end{eqnarray}
Substituting (\ref{3.17}) and (\ref{3.18}) into (\ref{3.15}) we get a  
quadratic equation for the product $\tilde{x}_1\tilde{x}_2$:
\begin{equation}
(\tilde{x}_1\tilde{x}_2)^2-\frac{2}{9}\left[(b-a)^2-(\beta -\alpha )^2\right]
(\tilde{x}_1\tilde{x}_2)-
\left(\frac{2}{9}\right)^2(b-a)^2(\beta -\alpha )^2=0.\label{3.19}
\end{equation}
The solution 
\begin{equation}
\tilde{x}_1\tilde{x}_2=\frac{2}{9}(b-a)^2\label{3.20}
\end{equation}
is just the Schwinger mass formula. It can be written in the form 
(\ref{0.13}). The widths of the particles do not contribute to the mass
formula. The second root 
$\tilde{x}_1\tilde{x}_2=-\frac{2}{9}(\beta -\alpha )^2$ is rejected as
negative. 

As $\tilde{x}_1,\tilde{x}_2$ and $(b-a)$ are positive, it follows from
(\ref{3.17}) and (\ref{3.18}) that $\tilde{y}_1,\tilde{y}_2$ and
$\beta -\alpha$ have the same sign.  So, $y_1$ and $y_2$ lie on
opposite sides of $y_8$. 

Multiplying (\ref{3.17}) by (\ref{3.18}) and using (\ref{3.20}), we find
the formula
\begin{equation}
\tilde{y}_1\tilde{y}_2=\frac{2}{9}(\beta -\alpha)^2\label{3.21}
\end{equation}
which is the analogue of the Schwinger mass formula.


Finally, we notice that (cf eq.~(\ref{0.21}))
\begin{equation}
\beta-\alpha=2(\kappa-\alpha) \label{3.22}
\end{equation}
and that WSR (\ref{3.21}) can be written in the form (cf (\ref{0.13}))
\begin{equation}
(\alpha -y_1)(\alpha -y_2)+2(\beta -y_1)(\beta -y_2)=0.\label{3.25}
\end{equation}

\subsection{Ideal nonet}
\label{sec:2.4}
If we use eq.~(\ref{0.7}) and the first three of eqs.~(\ref{0.8}) (for j=1,2,3),
we obtain eqs.~(\ref{0.17}) for complex masses 
\begin{equation}
z_1=a_c,\; \quad z_2=b_c,\;\quad{\rm and}\;\quad \mid{l_{1}}\mid^{2}=\frac{1}{3},\;
 \quad \mid{l_{2}}\mid^{2}=\frac{2}{3}.\label{3.26}
\end{equation}
Hence,
\begin{equation}
x_1=a,\;\quad x_2=b,\;\quad\Gamma _1=\Gamma _a,\;\quad\Gamma _2=\Gamma _b.
\label{3.27}
\end{equation}

Note that eq.~(\ref{0.7}) and three eqs.~(\ref{0.8}) give four real
conditions and three imaginary ones. The first two of the real
conditions determine $\mid{l_{1}}\mid^{2}$, $\mid{l_{2}}\mid^{2}$, the
remaining two give ideal values for $x_1$, $x_2$. For calculating
$\Gamma _1$, $\Gamma _2$ we have three imaginary equations. However,
for ideal masses the ideal $\Gamma _1$, $\Gamma _2$ follow from the
first two eqs.~(\ref{0.8}) and the third equation does not change the
result.

\section{The rectilineal stitch of the masses on the complex plane}
\label{sec:4}
The formulae (\ref{3.9}), (\ref{3.20}) and (\ref{3.26}) for the real
parts of the complex masses are identical with the corresponding formulae
for the real ones. So, the conditions of flavour-symmetry breaking
which operate in CMSR well reproduce data on the masses.

We assumed from the beginning that also in the WSR (\ref{3.10}),
(\ref{3.21}) and (\ref{3.26}) flavour-symmetry breaking factors are
correctly taken into account. However, this does not mean that they
have to describe well the real data. The particle widths depend also
on non-flavour factors which violate the WSR.  Let us call them for
brevity ``kinematical'', although they may include other effects.
Among ``kinematical'' factors the main role is played by strictly
kinematical ones -- the phase-space volume and conservation lows.
These factors may considerable disturb the widths of the low-mass
particles, but the higher is mass, the weaker is their influence due to
opening of new decay-channels. Therefore, violation of the WSR would be
significant in low-mass nonets and would weaken for more massive ones.
The data reveal such a tendency.  It is, therefore, likely that our
initial assumption is true (the flavour symmetry breaking is really
well described). Then violation of the WSR is a measure of the
``kinematical'' violation. However, it is difficult to evaluate on the
basis of WSR the size of the violation and attribute it to definite
particles of the nonet, because the sum rules do not include the
``triangulation point''. One can make the ``kinematical'' breaking
transparent combining WSR with the mass formula.

Eqs.~(\ref{3.17}) and (\ref{3.18}) connect the real and imaginary
parts of the masses of the S nonet. By means of eq.~(\ref{3.20}), we
obtain very simple result (cf (\ref{3.14})):

\begin{equation}
\frac{\tilde{y}_2}{\tilde{x}_2}=\frac{\tilde{y}_1}{\tilde{x}_1}
=\frac{\beta -\alpha }{b-a}=k_s.\label{3.28}
\end{equation}

{\em The points $(a,\alpha )$, $(K,\kappa )$,
  $(x_1,y_1)$, $(x_2,y_2)$, $(x_8,y_8)$, $(b,\beta )$ lie on a 
  straight line in the plane $(m^2,m\Gamma )$. Also the points 
  $(m_a,\Gamma _a)$, $(m_K,\Gamma _K)$ etc. lie
  on a straight line with the same slope in the plane $(m,\Gamma )$.
  The slope $k_s$ is indefinite.} 

So, the masses form a rectilineal stitch (RS) on the complex plane.

Eqs.~(\ref{3.28}) ultimately explain the nature of violation of the
WSR of the S nonet. If the slope of the stitch $k_s\neq{0}$ (true for
all of S nonets), then the WSR and the S mass formula are equivalent and
satisfy the same conditions of the broken flavour-symmetry. Therefore,
flavour breaking does not operate and violation of the WSR of the S
nonet is ``kinematical''.

Reversing the argument, one can say that rectilineality of the stitch
is a result of flavour-symmetry breaking which identically influence
the mass formula and the WSR.

\section{Ordering rules}
\label{sec:5}

The expressions (\ref{0.14}), (\ref{0.15}) derived for an S nonet do
not guarantee positivity of $\mid{l_1}\mid^2$, $\mid{l_2}\mid^2$. This
is required additionally.  The requirement introduces further
restrictions on the masses which take the form of an ordering rule.
There are two allowed mass orderings of the S nonet (cf \cite{3})
implying also two distinct inequalities between some masses and two
different ranges of the mixing angle $\theta ^{Sch}$:
\begin{eqnarray}
a<x_1<b<x_2; \;\quad 2K<x_1+x_2;\;\quad \theta ^{Sch}>\theta ^{id},\label{3.28a}\\
x_1<a<x_2<b;\;\quad 2K>x_1+x_2;\;\quad \theta ^ {Sch}<\theta ^{id},\label{3.29}
\end{eqnarray}
where $\theta ^{id}=35.26^\circ$ is the ideal mixing angle.  The mass
ordering rules follow from eqs.~(\ref{0.13}), (\ref{0.14}) and
(\ref{0.15})) under conditions $\mid{l_1}\mid^2>0$,
$\mid{l_2}\mid^2>0$. The inequalities for the masses follow from the
ordering and the relation $a+b=2K$.  The inequalities for $\theta
^{Sch}$ follow from the ordering and the equation
\begin{equation}
\tan^2{\theta ^{Sch}}=\frac{\mid{l_{1}}\mid^{2}}{\mid{l_{2}}\mid^{2}}.\label{3.30}
\end{equation}

To obtain the ordering rules for widths we combine
\begin{eqnarray}
y_8&=&\alpha +\frac{2}{3}(\beta -\alpha ),\label{3.31}\\
y_8&=&\beta -\frac{1}{3}(\beta -\alpha )\label{3.32}
\end{eqnarray}
with eqs.~(\ref{3.17}) and (\ref{3.18}) and observe that for the nonet
(\ref{3.28a})
\begin{equation}
\tilde {x}_1<\frac{2}{3}(b-a),\;\quad \tilde{x}_2>\frac{1}{3}(b-a)\label{3.33}
\end{equation}
and for the nonet (\ref{3.29})
\begin{equation}
\tilde{x}_1>\frac{2}{3}(b-a),\;\quad \tilde{x}_2<\frac{1}{3}(b-a).\label{3.34}
\end{equation}
We thus find two possible width orderings for each mass ordering: 
for the rule (\ref{3.28a}) we find
\begin{equation}
\alpha <y_1<\beta <y_2\;\quad {\rm or}\;\quad \alpha >y_1>\beta >y_2\label{3.35}
\end{equation}
and for the rule (\ref{3.29})
\begin{equation}
y_1<\alpha <y_2<\beta \quad {\rm or}\quad y_1>\alpha >y_2>\beta. \label{3.36}
\end{equation}

For the GMO nonet the only restriction on the masses (eq.~(\ref{3.11}))
follows from requirement of positivity of $\mid{l}\mid^2$s.
Therefore, besides the rules (\ref{3.28a}) and (\ref{3.29}) there are
also possible the inequalities
\begin{equation}
a<x_1<x_2<b \quad{\rm and} \quad x_1<a<b<x_2,\label{3.37}
\end{equation}
where the conditions (\ref{3.28a}), (\ref{3.29}) for $K$ and
$\theta^{GMO}$ do not hold.  In particular, the equality 
$\theta^{GMO}=\theta^{id} $ is possible for a nonideal nonet: 
$x_1\neq{a}$,  $x_2\neq{b}$. Therefore, for such a nonet the value of 
$\theta^{GMO}$ would not yield a criterion of ideality. However, we do 
not know such a nonet as yet.

\section{Bird's eye view on nonet data}
\label{sec:6}

Tab.~1 collects data on seven S nonets ordered by increasing $K$. To
make the data more transparent the physical masses and widths quoted
from PDG \cite{2} are supplemented with the calculated values of
$m_b=\sqrt{b}$ and $\Gamma_b=\frac{\beta}{m_b}$ (``mass'' and
``width'' of the ideal state $s\bar{s}$).  We also indicate for each
nonet the mixing angle $\theta{}$ as well as the mass and width ordering.

In the IM nonet the numbers from neighbouring columns 3 and 4 as well
as 5 and 6 would be equal. As they are not, the nonets are not ideal.
Instead, all of them are the S nonets. That can be checked by
saturating the Schwinger mass sum rule with the masses lying within
the bounds of experimental error. These masses define mixing angle
$\theta^{Sch}$. On the other hand, we can calculate $\theta^{GMO}$
using mean experimental values of the masses.  The mixing angle
$\theta$ (we assume $0<\theta<\frac{\pi}{2}$) quoted in Tab.~1 is in
most cases the $\theta^{GMO}$ one. The calculated $\theta{}$ often
have big errors. In several cases we cut the errors using restriction
(\ref{3.28a}) or (\ref{3.29}). Observe, that for the orderings
(\ref{3.28a}) or (\ref{3.29}) which are allowed by Schwinger mass
formula, the requirements formulated for $\theta^{Sch}$ are also valid
for $\theta^{GMO}$ and that these angles are not far removed from each
other in the vicinity of $\theta^{id}$.

Some further remarks are in order.

Two known pseudoscalar multiplets are not included in Tab.~1.
The first is the nonet $\pi$, $K$, $\eta$, $\eta{'}$ which is not the
S one and, besides, has no hadronic decays (except of $\eta{'}$).
The second is the multiplet (nonet?) $\pi(1300)$, $K(1460)$, 
$\eta(1295)$, $\eta(1440)$ for which we cannot establish even the mass
ordering, owing to big errors of $\pi(1300)$ and $K(1460)$ masses.

The masses of the unphysical states $K_A$ $(1^{++})$ and
$K_B$$(1^{+-})$ are required to satisfy the S nonet constraints. That
makes their mixing angles the $\theta^{Sch}$ ones by definition.  The
states $K_A$ and $K_B$ are superpositions of the physical states
$K_1(1270)$, $K_1(1400)$ and therefore the masses of $K_A$ and $K_B$
must satisfy an additional condition imposed by mixing. These three
constraints prove to be very restrictive, and we find that the values
of $K_A$ and $K_B$ obeying them are contained within narrow intervals
which are comparable with error ranges of the $K_1(1270)$ and $K_1(1400)$
masses. Also we find that values of $a_1$-meson mass allowed by these
constraints cover only part of the range of experimental data. For the
details of the procedure, see the Appendix.

The values of $ \beta{}$ (see eq.~(\ref{3.7a})) for the nonets
$1^{--}$ and $1^{++}$, calculated from the data on $a$- and $K$-meson
widths, come down deeply into the region of negative values (Tab.~1),
while the sum rules predict $\beta>y_2$. In both cases we accept the
width of $a$-meson as a measure of ``normal'' (unsuppressed) width
suitable for the nonet (even if the error of the $a_1$-meson width is
so big).  In the $1^{--}$ nonet, $\beta{}<0$ explicitly indicates a 
deficiency of $K^*$-meson width.  In the nonet $1^{++}$, the scope of
the calculated $\beta$ comprise negative values as well as positive ones.
The negative values may be explained by the big error of the $a_1$ width,
without invoking a deficit of the $K_A$ width. In that case (if the width
of $K_A$ were really not reduced) the $a_1$ width would be close to
the lower limit of the experimental value.  Similar remarks can also be 
made for other multiplets. However, that and other disagreements
between prediction and data are better seen from mass-width diagrams.


Fig.~1 exhibits mass-width diagrams of the same S nonets. On each
diagram, besides the points $(m,\Gamma)$ representing the observed
mesons, we draw the straight line crossing two or more of the points.
With one exception of $K_3(1780)$ (the data on $\eta_2(1870)$ we
consider uncertain), the line is drawn in such a way that there are no
experimental points lying above it. That follows from the guess that
deviations from the RS occur only downward. The guess itself reflects
observation that suppressing-decay mechanisms are well known and
frequent, while nothing certain is known about enhancing-decay
mechanisms. For some of the nonets we also show dominating channels of
the decay. We wonder whether the straight lines can be identified with
the RS.

The most striking feature of these diagrams is the negative slope of
the straight lines for all nonets (heavier particles have smaller
width).

Let us discuss the diagrams of some nonets in more detail.  Begin from
the nonets $1^{--}$, $2^{++}$ where we have the most complete data. We
draw the straight lines over the points ($\rho$,$\Phi$) for $1^{--}$
and ($f_2$,$f_2'$) for $2^{++}$. The dominating channels of the
hadronic decays of these particles $\rho \rightarrow \pi\pi$, $\Phi
\rightarrow K\bar{K}$, $f_2 \rightarrow \pi\pi$, $f_2' \rightarrow
K\bar{K}$ are J,I,P,C,G,S - allowed, but in the case of $\Phi
\rightarrow K\bar{K}$ the phase-space is small and the width would be
relatively reduced.  Therefore, if we want to identify these lines
with the RS, we should remember that $1^{--}$ line is steeper.  From
the Fig.~1 we read off $k_s(1^{--})$=-0.56 and $k_s(2^{++})$=-0.44.
    
For the nonet $3^{--}$ we draw the line over the points $\omega_3$,
$\rho_3$, $\Phi{_3}$. Each of the particles has a phase-space large
enough for many decay channels.  Apparently, at these energies the
number of opened channels is sufficient for the particles to satisfy
the RS equation.  From the Fig.~1 we read off $k_s(3^{--})$=-0.43.

\pagebreak 
It is, therefore, likely that the slopes of RS are similar
for the nonets $1^{--}$, $2^{++}$, $3^{--}$ and are 
concentrated inside the interval $k_s=-0.5\div -0.4$.

The diagrams of $2^{-+}$ and $0^{++}$ nonets do not conflict with this
observation.  We do not appeal to them, as they suffer from uncertain
data on the isoscalar mesons ($2^{-+}$) or from an uncertain nonet
assignment ($0^{++}$).

The situation is different with $1^+$ nonets. For the mesons of both
$1^{+-}$ and $1^{++}$ nonets the decays into $\pi\pi$ and $K\bar{K}$
are kinematically forbidden. Their two-particle decays producing
heavy-meson ($\rho{}, \omega$, etc.) and many-particle decays are
mostly more or less suppressed by the phase space.  For the $1^{++}$
nonet where we have more data, the straight line shown on Fig.1
crosses the points $a_1$ and $f_1(1420)$.  Perhaps the decays of the
$a_1$-meson ($a_1 \rightarrow (\pi\rho)_{S wave}$ etc.) may be
considered unsuppressed, as this particle has many decay channels
opened and a huge width (although the error is extremally big); but
the decays $f_1(1420) \rightarrow K\bar{K}\pi$,
$K\bar{K}^*(892)+c.c.$ are clearly suppressed. So the line is based on
the particle which has kinematically suppressed decays. Therefore, we
cannot accept it as the RS one. A similar situation holds for the
$1^{+-}$ nonet.



\begin{table}
\begin{center}
\caption{Schwinger nonets of mesons ($m$ and $\Gamma$ in MeV).
Nonets, ordered by increasing $m_K$, are described in three rows 
containing: masses; widths; mixing angle, mass and width ordering. 
Subscripts $a$, $K$, $1$, $2$ denote isotriplet, isodublet and 
isoscalar states.  $m_b$=$\sqrt{b}$ and $\Gamma_b$=$\frac{\beta}{m_b}$ 
are calculated. In ordering rules $a$, $b$, $x_1$, $x_2$ are  masses 
squared and  $\alpha{}$, $\beta{}$, $y_1$, $y_2$ are products of mass 
and width.
Mixing angle $\theta$ is the $\theta^{GMO}$, except of $1^{++}$, $1^{+-}$ 
where it is $\theta^{Sch}$. $K_A$, $K_B$ are unphysical nonet states 
(Appendix). Notations and data quoted from RPP  \cite{2}.}

\label{tab1}
\begin{center}
\small
  \addtolength{\topmargin}{-25pt} \addtolength{\textheight}{25pt}
\begin{tabular}{|r@{}l||c||c|c||c|c|}
\hline
&&$m_K$&$m_a$&$m_1$&$m_b$&$m_2$\\
\multicolumn{2}{|c||}{$J^{PC}$}&$\Gamma_K$&$\Gamma_a$&$\Gamma_1$&$\Gamma_b$
&$\Gamma_2$\\
\hline
\multicolumn{2}{|c||}{particles}&\multicolumn{1}{c}{$\theta^{GMO}$}&
\multicolumn{2}{c}{mass ordering}&\multicolumn{2}{c|}{width ordering}\\
\hline\hline
\multicolumn{2}{|c||}{$1^{--}$}&$893.88\pm 0.26$&$769.3\pm 0.8$&$782.57\pm 0.12$&
$1001.7\pm 1.1$&$1019.456\pm 0.020$\\
\cline{1-2}
$\bullet$&$\rho(770)$&&&&&\\
$\bullet$&$K^*(892)$&$50.7\pm 0.8$&$149.2\pm 0.7$&$8.44\pm 0.09$&
$-24.8\pm 2.1$&$4.26\pm 0.05$\\
$\bullet$&$\omega(782)$&&&&&\\
\cline{3-7}
$\bullet$&$\Phi(1020)$&\multicolumn{1}{c}{$(39.28\pm 0.16)^\circ$}&
\multicolumn{2}{c}{$a<x_1<b<x_2$}&
\multicolumn{2}{c|}{$\alpha>y_1>\beta>y_2$}\\
\hline\hline
\multicolumn{2}{|c||}{$1^{+-}$}&$1324\pm 8$&$1229.5\pm 3.2$&$1170\pm 20$&
$1414\pm 9$&$1386\pm 19$\\
\cline{1-2}
$\bullet$&$b_1(1235)$&&&&&\\
$\bullet$&$K_B$&$135\pm 17$&$142\pm 9$&$360\pm 40$&
$130\pm 40$&$91\pm30$\\
$\bullet$&$h_1(1170)$&&&&&\\
\cline{3-7}
$\bullet$&$h_1(1380)$&\multicolumn{1}{c}{$0\div 35.26^\circ$}&
\multicolumn{2}{c}{$x_1<a<x_2<b$}&
\multicolumn{2}{c|}{$y_1>\alpha>y_2>\beta$}\\
\hline\hline
\multicolumn{2}{|c||}{$1^{++}$}&$1340\pm 8$&$1230\pm 40$&$1281.9\pm 0.6$&
$1420\pm 12$&$1426.3\pm 1.1$\\
\cline{1-2}
$\bullet$&$a_1(1260)$&&&&&\\
$\bullet$&$K_A$&$134\pm 16$&$250\div 600$&$24.0\pm 1.2$&
$-447\div 89$&$55.5\pm 2.9$\\
$\bullet$&$f_1(1285)$&&&&&\\
\cline{3-7}
$\bullet$&$f_1(1420)$&\multicolumn{1}{c}{$35.26^\circ \div 41.00^\circ$}&
\multicolumn{2}{c}{$a<x_1<b<x_2$}&
\multicolumn{2}{c|}{$\alpha>y_1>\beta>y_2$}\\
\hline\hline
\multicolumn{2}{|c||}{$0^{++}$}&$1412\pm 6$&$984.7\pm 1.2$&$980\pm 10$&
$1737\pm 11$&$1713\pm 6$\\
\cline{1-2}
$\bullet$&$a_0(980)$&&&&&\\
$\bullet$&$K_0(1430)$&$294\pm 23$&$50\div 100$&$40\div 100$&
$380\div 490$&$125\pm 10$\\
$\bullet$&$f_0(980)$&&&&&\\
\cline{3-7}
$\bullet$&$f_0(1710)$&\multicolumn{1}{c}{$(33.5\pm 2.0)^\circ$}&
\multicolumn{2}{c}{$x_1<a<x_2<b$}&
\multicolumn{2}{c|}{$y_1>\alpha>y_2>\beta  $}\\
\hline\hline
\multicolumn{2}{|c||}{$2^{++}$}&$1429.0\pm 1.4$&$1318.0\pm 0.6$&$1275.4\pm 1.2$&
$1532.0\pm 3.1$&$1525\pm 5$\\
\cline{1-2}
$\bullet$&$a_2(1320)$&&&&&\\
$\bullet$&$K_2^*(1430)$&$103.8\pm 4.0$&$107\pm 5$&$185.1^{+3.4}_{-2.6}$&
$101.5\pm 11.9$&$76\pm 10$\\
$\bullet$&$f_2(1270)$&&&&&\\
\cline{3-7}
$\bullet$&$f'_2(1525)$&\multicolumn{1}{c}{$(30.67^{+1.56}_{-1.72})^\circ$}&
\multicolumn{2}{c}{$x_1<a<x_2<b$}&
\multicolumn{2}{c|}{$y_1>\alpha>y_2>\beta$}\\
\hline\hline
\multicolumn{2}{|c||}{$2^{-+}$}&$1773\pm 8$&$1670\pm 20$&$1617\pm 5$&
$1870\pm 33$&$1842\pm 8$\\
\cline{1-2}
$\bullet$&$\pi_2(1670)$&&&&&\\
$\bullet$&$K_2(1770)$&$186\pm 14$&$259\pm 10$&$181\pm 11$&
$122\pm 49$&$225\pm 14$\\
&$\eta_2(1645)$&&&&&\\
\cline{3-7}
&$\eta_2(1870)$&\multicolumn{1}{c}{$0\div 35.26^\circ$}&
\multicolumn{2}{c}{$x_1<a<x_2<b$}&
\multicolumn{2}{c|}{$y_1>\alpha>y_2>\beta$}\\
\hline\hline
\multicolumn{2}{|c||}{$3^{--}$}&$1776\pm 7$&$1691\pm 5$&$1667\pm 4$&
$1857\pm 11$&$1854\pm 7$\\
\cline{1-2}
$\bullet$&$\rho_3(1690)$&&&&&\\
$\bullet$&$K_3^*(1780)$&$159\pm 21$&$161\pm 10$&$168\pm 10$&
$158\pm 53$&$87^{+28}_{-23}$\\
$\bullet$&$\omega_3(1670)$&&&&&\\
\cline{3-7}
$\bullet$&$\Phi_3(1850)$&\multicolumn{1}{c}{$(32.0^{+3,3}_{-7.5})^\circ$}&
\multicolumn{2}{c}{$x_1<a<x_2<b$}&
\multicolumn{2}{c|}{$y_1>\alpha>y_2>\beta$}\\
\hline
\end{tabular}
\end{center}
\end{center}
\end{table}


\begin{figure}
\includegraphics[height=17cm]{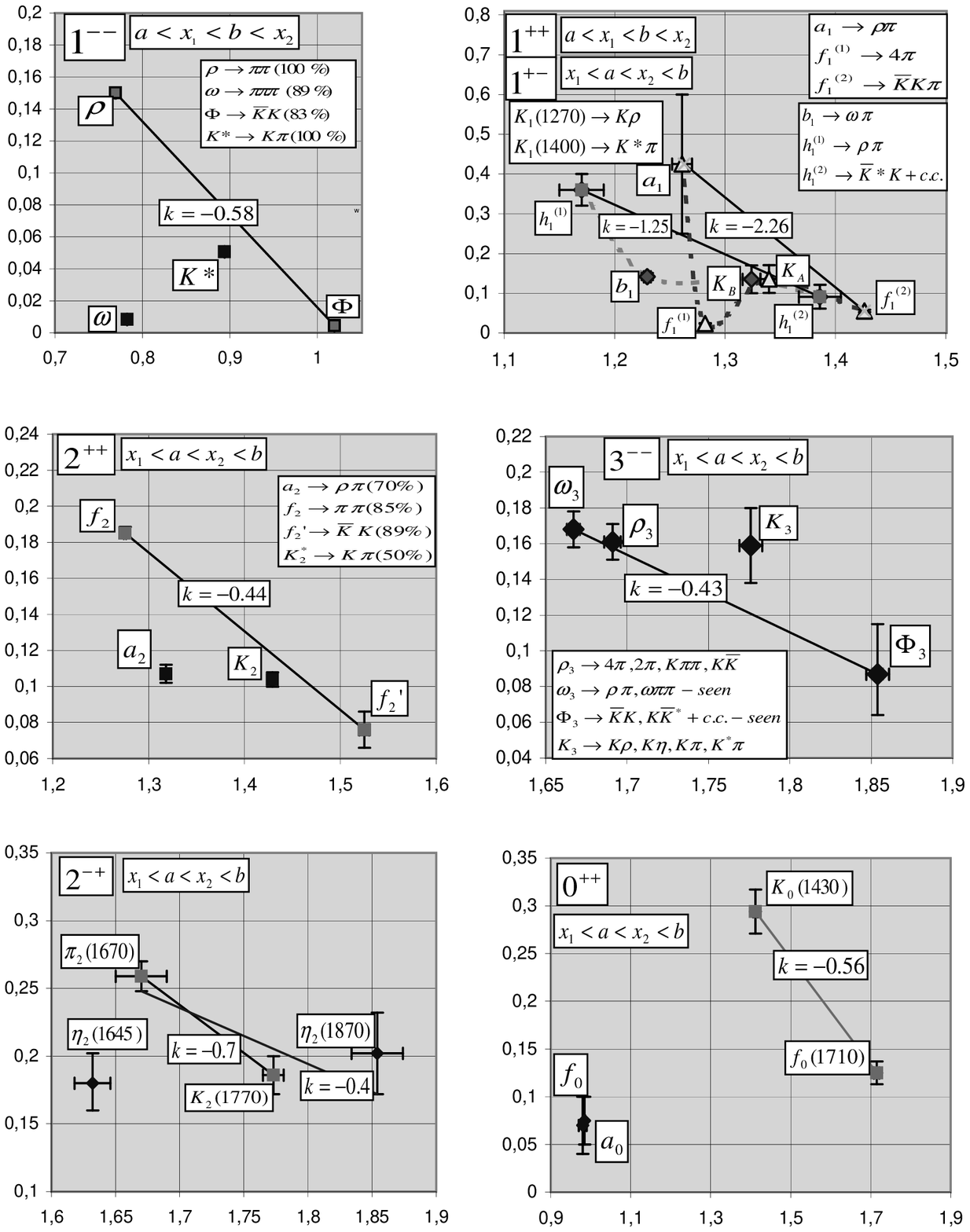}
\caption{Mass-width diagrams of the S nonets. On the axes m and $\Gamma$ in GeV. Shown dominating decays of some particles. Full straight lines crossing two or more points are supposed to be nearest by RS of the nonet. For $3^{--}$ and $2^{++}$ they are expected to lie along the RS. For $1^{--}$ small phase-space pushes down the observed point $\Phi{}$. For $1^+$ nonets the straight lines cannot be identified with RS.}
\label{fig.1.}
\end{figure}

\section{Degenerate octet}
\label{sec:5}
The ninth meson does not mix with the octet, if $l_{2}^{2}=0$. Then,
$x_1=x_8$, $l_{1}^{2}=1$ and $x_2$ is arbitrary. It follows also from
(\ref{3.20}) that $b=a$ and thus $K=a=x_8$. Then, from (\ref{3.17}) we
find $\tilde{y}_1=0$, from (\ref{3.21}) $\beta =\alpha$ and thus
$\beta =\kappa =y_8$. We see that the restrictions put by the first two
exotic commutators (\ref{0.1}) on the octet states provide their full
degeneration. This result remains unchanged, if further exotic
commutators are included.

Tab.~2 shows two $1^{--}$ multiplets and the $4^{++}$ multiplet which
can be understood as degenerate octets. The degeneracy of these octets
is clearly seen from the data on the masses. It is confirmed by data on
the widths.

On the other hand the data on the widths confirm the suggestion 
that at these energies the WSR are well satisfied.

\begin{table}
\caption{Degenerate octets ($m$ and $\Gamma{}$ in MeV \cite{2})}
\label{table2}
\begin{center}\small
\begin{tabular}{|r@{}l||c|c|c|}
\hline
&&$m_a$&$m_K$&$m_1$\\
\multicolumn{2}{|c||}{$J^{PC}$}&$\Gamma_a$&$\Gamma_K$&$\Gamma_1$\\
\hline

&$1^{--}$&&&\\
$\bullet$&$\rho(1450)$&$1465\pm 25$&$1414\pm 15$&$1419\pm 31$\\
$\bullet$&$K^*(1410)$&&&\\
$\bullet$&$\omega(1420)$&$310\pm 60$&$232\pm 21$&$174\pm 60$\\
\hline
&$1^{--}$&&&\\
$\bullet$&$\rho(1700)$&$1700\pm 20$&$1717\pm 27$&$1649\pm 24$, \quad$1680\pm 20$\\
$\bullet$&$K^*(1680)$&&&\\
$\bullet$&$\omega(1650)$&$240\pm 60$&$322\pm 110$&$220\pm 35$, \quad $150\pm50$\\
$\bullet$&$\Phi(1680)$&&&\\
\hline&$4^{++}$&&&\\
$\bullet$&$a_4(2040)$&$2011\pm 13$&$2045\pm 9$&$2025\pm 8$\\
$\bullet$&$K_4(2045)$&&&\\
$\bullet$&$f_4(2050)$&$360\pm 40$&$198\pm 30$&$194\pm13$\\
\hline
\end{tabular}
\end{center}
\end{table}

\section {Summary and discussion}
\label{sec:6}

The descripton of unstable particle quantum states via attributing to
them a complex mass is under study from various points of view for a
long time (see e.g., \cite{2a} and references therein). We make an
attempt, for the first time, to apply the idea of a complex mass for
the derivation of extended mass formulae within the early proposed
algebraic approach \cite{1} based on exotic commutators of the
$SU(3)_f$ charge operators and their time derivatives in the infinite
momentum frame system. The new sum rules including hadronic widths of
the resonances are compared with the data for ten meson multiplets and the
gains and problems encountered are of a certain interest and may also
serve as a starting point for a further investigation.

The real parts of the complex mass sum rules (CMSR) give relations
between the real parts of the complex mass squared which are identical
with the well known mass formulae of the broken flavour symmetry for
real masses.
 
The CMSR predict two possible kinds of octet: Gell-Mann -- Okubo
(with masses satisfying the GMO mass formula) and degenerate. So, the
latter is obtained not only for exact symmetry but also under the CMSR
constraints. The CMSR predict width degeneration of the mass
degenerate octet particles. The data shown in Tab.~2 demonstrate that
two $1^{--}$ multiplets and the $4^{++}$ one are degenerate octets.
The degeneration is seen not only from the masses, but also from the
widths.

The CMSR predict three kinds of nonets: Gell-Mann -- Okubo, Schwinger
(S) and ideally mixed (IM). All the observed nonets having hadronic
widths are S ones.  As the S nonet well describes the masses, we
may think that it correctly takes into account the flavour breaking
factors.

The S nonet mesons are submitted to definite mass ordering.  There are
two allowed orderings: $x_1<a<x_2<b$ and $a<x_1<b<x_2$ (for the
notations, see Sect.~2).  The mass ordering decides whether the mixing
angle of the nonet is smaller or bigger than $\theta^{id}$. It also
decides whether $2K$ is bigger or smaller than $x_1+x_2$. Thus the
mass ordering is a relation characterizing nonet; likewise the mixing
angle characterize it.

The imaginary parts of the CMSR give relations between the imaginary
parts of the complex mass squared. These relations -- the width sum
rules (WSR) of GMO, S and IM types -- connect the total widths of the
nonet mesons. WSR have shapes identical with the mass formulae, but,
in contrast, they are not satisfied by data in general. The reason is
an extra reduction of the widths of the individual particles by
``kinematical'' factors (we mean not only strictly kinematical
factors, like phase-space and E,J,I,S,P,C,G - conservation, but also
all other unflavour ones).  This violation diminishes along with
growth of the mean nonet mass and disappears at about 1.5~GeV.
 
In the S nonets the ordering rules are in force also for the imaginary
parts of the masses. They are correlated with the mass ordering rules,
but also are not satisfied by the observed widths of the low-mass
nonets.

Tab.~1 collects seven S nonets. For each of them are shown masses, 
widths, mass and width ordering, and mixing angle. Deviations 
of the isoscalar mesons from ideal values are apparently seen.

For the S nonet CMSR predict a linear relation between the masses and
the widths of all particles. Hence, the complex masses form a
rectilineal stitch (RS) on the complex plane with a slope $k_s$
depending of the nonet. As $k_s\neq 0$, the Schwinger WSR is
equivalent to the S mass formula and complies with the conditions
breaking flavour symmetry. Therefore, violation of the WSR cannot be a
result of the flavour symmetry breaking and consequently {\em
  violation of the WSR for the S nonet is ``kinematical''.}
 
To construct the RS of a nonet we need the particles decaying in the
``kinematically'' unsuppressed processes. We can do that for a nonet
having masses above 1.5~GeV where many channels of the decay are
oppened and WSR are satisfied. $3^{--}$ is such a nonet.  Also for a
low mass nonet the RS can be determined approximately, if there exist
two particles with ``kinematically'' unsuppressed dominating decays.
The nonets $1^{--}$ and $2^{++}$ are examples. Fig.~1 demonstrates
mass-width diagrams of seven nonets. For the nonets $1^{--}$ and
$2^{++}$ the RS is constructed as the highest lying straight line
crossing two experimental points $(m,\Gamma)$ of the diagram. For the
$1^+$ nonets such a line cannot be identified with the RS, because of
lack of two particles decaying in ``kinematically'' unsuppressed
reaction.

In the cases where $k_s$ is known we can evaluate the ``kinematical''
suppression of a particle as the difference between predicted (lying
on RS) and observed value of the width.
 
The rectilineality of the stitch and the stitch itself result from
flavour-symmetry breaking (for an exactly symmetric multiplet there
would be only one point), but the slope of the stitch $k_s$ is
undetermined.  Data show that $k_s$ is negative for all nonets. They
also suggest that the slopes are not much different from each other
and are concentrated somewhere in the region $k_s\approx-0.5\div
-0.4$. The origin of such behaviour of $\Gamma(m)$ is unknown. We can
only remark that it resembles the behaviour of the strong coupling
constant $\alpha{_s}$, also decreasing with icreasing mass and being
flavour independent.

Let us go back to the idea with which we began this paper.  Hadron
widths do not influence the mass formula of broken flavour symmetry
and the mixing angle of the nonet. Therefore, one could once discover
them and describe the nonet. That is the reason why hadron
spectroscopy could completely ignore the data on the total widths of
the particles. However, hadrons do have finite widths and hadron
spectroscopy should obligatorily describe them. Moreover, the widths
are considerable and may be important for description of the multiplet
as a whole, not only as attributes of individual particles. But where
is a trace of that?  CMSR introduce the widths into our scope. They
predict not only the WSR connecting the total widths of the nonet
mesons, but also the RS being the result of the interplay between real
and imaginary parts of the CMSR.  That is the place where we can
expect something new. The RS is such a relation characterizing all S
nonets; perhaps the slope $k_s$ characterizes the nonet individually.


Much attention is devoted nowadays to the existence of glueballs.
Search for these states requires detailed information on the
multiplets. Glueball cannot be discovered by analysing the properties
of a single particle. Even more, such a discovery would not be
convincing.  The glueball state with nonexotic quantum numbers $J^{PC}$
should mix with isoscalar $q\bar{q}$ states.  Therefore, the way to
identify the glueball is to look for meson decouplets including three
isoscalar physical mesons and investigate them \cite{3}. Sum rules for
complex masses may be useful in such analysis. Investigating of the
sum rules for decouplets is in progress.

\section{Acknowledgments}
The author thanks Profs. S.B.Gerasimov, P.Kosi\'nski, V.A.Meshcheryakov for
valuable discussions and Prof. W.Tkaczyk, Dr K.Smoli\'nski, Dr
J.Olejniczak for help in computer operations.

\appendix
\section*{Appendix: $1^+$ multiplets as S nonets}
Below, $1^{++}$ mesons are called $a_A$, $K_A$, $x_{A1}$, $x_{A2}$
(nonet A) and $1^{+-}$ mesons are called $a_B$, $K_B$, $x_{B1}$,
$x_{B2}$ (nonet B) (cf Tab.~1). We assume that each of the nonets
satisfies the Schwinger mass formula. As a function of $a$ and $K$
this formula is the equation of ellipse:
\begin{equation}
3a^2+8K^2-8aK+ a(x_1+x_2)-4K(x_1+x_2)+3x_1x_2=0.\label{A.1}
\end{equation}
The parameters of the ellipse are determined by the masses of the isoscalar
mesons $x_1$, $x_2$ which fix the position of the ellipse centre:
\begin{equation}
a^{Cr}=K^{Cr}=\frac{1}{2}(x_1+x_2),\label{A.1a}
\end{equation}
and the magnitude of its axes $\bf{a}$,$\bf{b}$ being proportional to
$(x_2-x_1)$.  However, they do not influence the axes' ratio
($\bf{a}/\bf{b}$=$3.6$) and the orientation of the ellipse in the plane
(the angle beween big axis and obscissa is $29^\circ$). So these
quantities are the same for all S nonets.

As the straight line $2K$=$x_1+x_2$ crosses the centre of the ellipse, the
mass ordering (\ref{3.28a}) or (\ref{3.29}) decides whether $K$ lies
below or above its diameter.  

The physical states of the $1^+$ mesons $K_1(1270)$ and $K_1(1400)$
are mixed states of the $K_A$ and $K_B$:
\begin{eqnarray}
|K_1(1270)\rangle&=&|K_A\rangle\cos\Phi+|K_B\rangle\sin\Phi,\nonumber\\
|K_1(1400)\rangle&=&-|K_A\rangle\sin\Phi+|K_B\rangle\cos\Phi,\label{A.2}
\end{eqnarray}
where $\Phi$ is the mixing angle \cite{2}. For the masses squared of these
mixed states we have
\begin{equation} 
K_A+K_B=K_1(1270)+K_1(1400).\label{A.3}
\end{equation}
We are looking for such values of $K_A$ and $K_B$ which satisfy this
equation and eq.~\ref{A.1}) for each nonet. 
  
{\em Ellipse A}

The masses of the $1^{++}$ isoscalar mesons $f_1(1285)$ and
$f_1(1420)$ have negligible errors, so the ellipse A is precisely
determined. It is shown on Fig.~2.  We can see that $a_A$ may be
assigned to the S nonet, only if
 \begin{equation}
 a_A>(1225 MeV)^2.\label{A.4}
\end{equation}
This, together with the experimental limit \cite{2}, gives
\begin{equation}
(1225 MeV)^2<a_A<(1270 MeV)^2. \label{A.5}
\end{equation}
From Fig.~2 we find
\begin{equation}
(1277 MeV)^2<K_A<(1347 MeV)^2.\label{A.6}
\end{equation}

\begin{figure}
  \centering
  \includegraphics{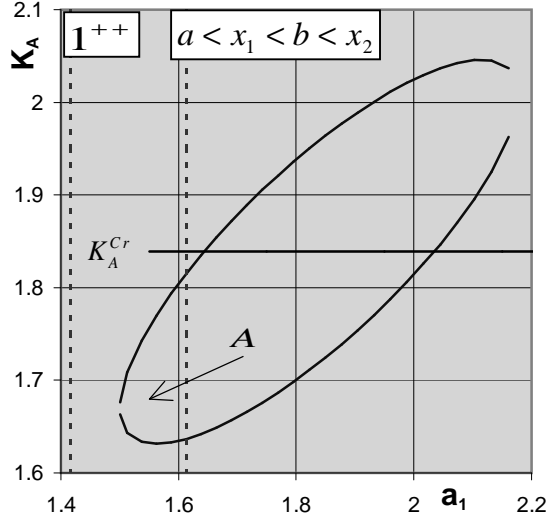}
\caption{Ellipse A. Schwinger mass formula as function of  $a_1$ and $K_A$. 
Dotted lines bound error corridor of $a_1$. Allowed $K_A$ belong to segment 
$A$ of the ellipse. They lie below diameter $K_A^{Cr}$ as required 
by mass ordering. On the axes $a_1$ and $K_A$ in $GeV^2$.}
\label{fig.A.1}
\end{figure}

{\em Ellipse B}

The mases of the $1^{+-}$ isoscalar mesons $h_1(1190)$, $h_1(1380)$ have
considerable errors which influence the parameters of the ellipse
(\ref{A.1}).
\begin{figure}
\centering
\includegraphics{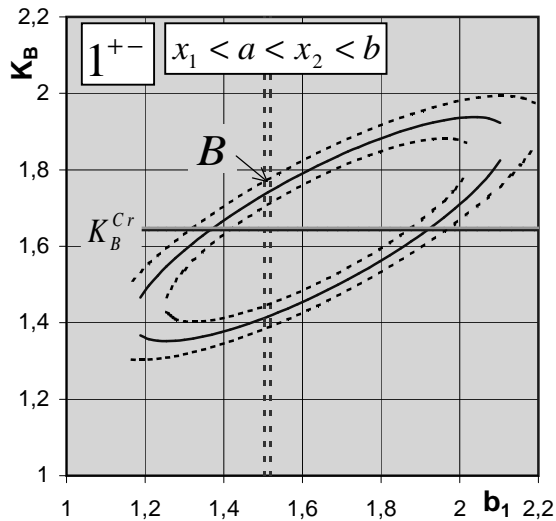}
 \caption{Ellipse B. Mass errors of $h_1^{(1)},h_1^{(2)}$ create 
corridor shown between dotted ellipses. Its crossing with $b_1$ corridor 
produces  solutions  for $K_B$ above and  below ellipse diameter $K_B^{Cr}$. 
Choice $K_B>K_B^{Cr}$ follows from mass ordering. On the axes $b_1$ and 
$K_B$ in $GeV^2$.}
\label{fig.A.2}
\end{figure}
The Fig. 3 presents three ellipses corresponding to the mean
experimental values of $x_{B1}$, $x_{B2}$ and to these two
combinations of their lower and uper experimental limits which give
the ellipses with minimal and maximal axes.  Putting for $a_B$ its
experimental value, we find two regions of solutions for $K_B$.  The
lower solutions are rejected by the mass ordering (\ref{3.29}) and we
obtain
\begin{equation} 
(1307MeV)^2<K_B<(1332MeV)^2.\label{A.7}
\end{equation}
Observe that in spite of the less accurate data on the isoscalar meson
masses, the evaluation of the $K_B$ is more accurate than of the
$K_A$ one.

Summing (\ref{A.6}) and (\ref{A.7}) we find
\begin{equation}
(1828MeV)^2<K_A+K_B<(1895MeV)^2.\label{A.8}
\end{equation}
On the other hand, from the data on the masses of the $K_1$ mesons we
have
\begin{equation}
(1884MeV)^2<K_1(1270)+K_1(1400)<(1904MeV)^2.\label{A.9}
\end{equation}
Comparing (\ref{A.9}) with (\ref{A.7}), we conclude that
\begin{equation}
K_B<\frac{1}{2}(K_1(1270)+K_1(1400)),\label{A.10}
\end{equation}
and therefore,
\begin{equation}
K_B<K_A.\label{A.11}
\end{equation}
Comparing (\ref{A.8}) with (\ref{A.9}), we find that the restrictions
on the $K_A$- and $K_B$-meson masses following from the nonet assignment
are compatible with the observed mass values of $K_1$-mesons within
the narrow interval of the sum $K_A$+$K_B$:
\begin{equation}
(1884MeV)^2<K_A+K_B<(1895MeV)^2,\label{A.12}
\end{equation}
where the upper limit of the sum is the sum of the individual upper 
limits of $K_A$ (\ref{A.6}) and $K_B$ (\ref{A.7}). Therefore, it
immediately folows that adjustable values of $K_A$ and $K_B$ are:
\begin{eqnarray}
K_A&=&(1340\pm8)^2MeV^2,\label{A.13}\\
K_B&=&(1324\pm8)^2MeV^2.\label{A.14}
\end{eqnarray}
Returning to the ellipse A, we find the adjustable value of $a_A$:
\begin{equation}
 a_A=(1261\pm9)^2MeV^2.\label{A.15} 
\end{equation}

The mixing angle $\Phi$, which can be calculated from (\ref{A.13}),
(\ref{A.14}) and the physical masses of $K_1$ mesons, is charged with
a big error exceeding the difference $(\Phi^{mean}-45^\circ)$.
Therefore, for evaluating $\Gamma_A$ and $\Gamma_B$ (using the
physical $\Gamma_K$s) we put $\Phi$=$45^\circ$. The calculated widths
are indicated in Fig.~1.





\end{document}